\documentclass[12pt]{iopart}
\usepackage{graphicx}
\usepackage{epsfig}
\usepackage{ulem}
\usepackage{morefloats}
\usepackage{rotating}
\usepackage{color,ulem}
\usepackage{comment}
\usepackage{lineno}
\usepackage{amssymb}\usepackage{ulem}

\def\beq{\begin{equation}}
\def\eeq{\end{equation}}
\def\beqa{\begin{eqnarray}}
\def\eeqa{\end{eqnarray}}
\def\ban{\begin{eqnarray*}}
\def\ean{\end{eqnarray*}}
\def\bi{\begin{itemize}}
\def\ei{\end{itemize}}

\begin{document}

\title{Isoscaling in Dilute Warm Nuclear Systems}
\author{
Alex Rebillard-Souli\'e$^1$,
R\'emi Bougault$^1$,
Helena Pais$^2$,
Bernard Borderie$^3$,
Abdelouahad Chbihi$^4$,
Caterina Ciampi$^4$,
Quentin Fable$^5$,
John Frankland$^4$,
Emmanuelle Galichet$^{3,6}$,
Tom G\'enard$^4$,
Di\'ego Gruyer$^1$,
Nicolas Le Neindre$^1$,
Ivano Lombardo$^{7,8}$,
Olivier Lopez$^1$,
Loredana Manduci$^9$, 
Marian P\^arlog$^1$,
Giuseppe Verde$^{7,5}$
}
\address{
$^1$Universit\'e de Caen Normandie, ENSICAEN, CNRS/IN2P3, LPC Caen UMR6534, F-14000 Caen, France.\\
$^2$CFisUC, Department of Physics, University of Coimbra,
  3004-516 Coimbra, Portugal.\\
$^3$Universit\'e Paris-Saclay, CNRS/IN2P3, IJCLab, 91405 Orsay, France.\\
$^4$Grand Acc\'el\'erateur National d'Ions Lourds (GANIL),
CEA/DRF-CNRS/IN2P3, Boulevard Henri Becquerel, F-14076 Caen, France.\\
$^5$Laboratoire des 2 Infinis - Toulouse (L2IT-IN2P3),
Universit\'e de Toulouse, CNRS, UPS, F-31062 Toulouse Cedex 9, France.\\
$^6$Conservatoire National des Arts et M\'etiers, F-75141 Paris Cedex 03, France.\\
$^7$INFN - Sezione di Catania, 95123 Catania, Italy\\
$^8$Dipartimento di Fisica e Astronomia, Universit\`a di Catania, via S. Sofia 64, 95123 Catania, Italy.\\
$^9$Ecole des Applications Militaires de l'Energie Atomique, B.P. 19, F-50115 Cherbourg, France.
}
\eads{\mailto{rebillard@lpccaen.in2p3.fr}, \mailto{bougault@lpccaen.in2p3.fr}, \mailto{hpais@uc.pt}}

\begin{abstract}
Heavy-ion collisions are a good tool to explore hot nuclear matter below saturation density, $\rho_{0}$. It has been established that if a nuclear system reaches the thermal and chemical equilibrium, this leads to scaling properties in the isotope production when comparing two systems which differ in proton fraction. This article presents a study of the isoscaling properties of an expanding gas source exploring different thermodynamic states (density, temperature, proton fraction). This experimental work highlights the existence of an isoscaling relationship for hydrogen and $^{3}$He, $^{4}$He helium isotopes which agrees with the hypothesis of thermal and chemical equilibrium. Moreover, this work reveals the limitations of isoscaling when the two systems differ slightly in total mass and temperature. Also, a discrepancy has been observed for the $^{6}$He isotope, which could be explained by finite size effects or by the specific halo nature of this cluster.
\end{abstract}

\submitto{\jpg}
\maketitle
 
\ack{This work was partly supported by the PHC PESSOA Project 47833UB (France) and the FCT PESSOA Project No. 2021.09262.CBM (Portugal).
This work is based on an experiment performed at GANIL.}
\section{Introduction}
Nuclear isoscaling refers to the scaling properties in the isotope production when comparing two nuclear systems that differ in the neutron to proton ratio \cite{TsangPRC64bis}.
These properties give information about the neutron to proton degree of freedom and its equilibration.
Isoscaling was first observed in experimental data \cite{XuPRL85} and over a variety of de-excitation mechanisms like evaporation,
strongly damped binary collisions, and multifragmentation \cite{TsangPRL86}.
It has also been observed in a variety of statistical models \cite{TsangEur2006}.
Conditions for isoscaling in nuclear reactions were studied in \cite{TsangPRC64}.
\par
The symmetry-term coefficient of the
nuclear mass formula, C$_{sym}$, is directly correlated with isoscaling parameters \cite{TsangPRC64bis} \cite{BotvinaPRC65} \cite{OnoPRC68}. This relation, however, is only valid in the zero-temperature limit \cite{BotvinaPRC65}.
For finite temperature events, the exact C$_{sym}$ value extracted with the help of isoscaling is an apparent value which differs from the symmetry-term coefficient.
Nevertheless for T$>$5.0 MeV the isoscaling parameters remain strongly correlated with 
C$_{sym}$ and temperature through a function that is as yet unknown \cite{SouzaPRC80}.
Therefore, observation of isoscaling remains a test of equilibrium and further explorations are necessary to better understand the consequences of this law.
\par
Recently \cite{PaisPRL125,PaisJPG47} the INDRA collaboration presented new sets of data, identified as an expanding gas of light clusters ($^2$H, $^3$H,  $^3$He, $^4$He and $^6$He) and free nucleons.
Chemical equilibrium constants, density, temperature and total number of protons and neutrons for each emission time of the expanding gas were evaluated from the data and compared to the relativistic mean field (RMF) model of Ref.~\cite{PaisPRC97}.
It was shown that the suppression of the cluster concentration corresponds to important in-medium modifications of the binding energies.  
\par
We therefore have data sets whose thermodynamic state is known: temperature, density, total number of protons and neutrons. This allows a study concerning nuclear isoscaling when in-medium effects have been identified. 
The aim of the present article is to compare the data with the isoscaling law in a quantitative way, indicating the quality of the performed fits and explaining why, under certain conditions, the law is not strictly respected.
\section{Experimental details and thermodynamical landscape}
This section begins by describing the details of data taking and event selection. The mean characteristics (temperature, total mass, total proton fraction, baryon density) of the expanding gas source for the four studied reactions are then given.
\subsection{Experimental details}
The $4\pi$ multi-detector INDRA 
\cite{Pouthas1995} was used to study four reactions with beams of $^{136}$Xe and $^{124}$Xe, 
accelerated at 32 MeV/nucleon, and thin (530 $\mu$g/cm$^{2}$) targets of $^{124}$Sn and $^{112}$Sn. 
INDRA is a charged product multidetector,
composed of 336 detection cells arranged in 17 rings centered
on the beam axis and covering 90\% of the solid angle. 
INDRA can identify in charge fragments from Hydrogen to Uranium and in mass light fragments with low thresholds.
Two reactions
($^{124}$Xe+$^{124}$Sn and  $^{136}$Xe+$^{112}$Sn) 
were chosen to study chemical equilibrium hypothesis
since their projectile+target combined systems are identical.
\par
Several publications have been produced based on the data obtained from this experiment. A first article \cite{BougaultPRC97} concerning light charged particle production as a function of impact parameter has been published. 
In the following articles \cite{PaisPRL125,PaisJPG47,BougaultJPG19}, the most violent events were selected in order to study expanding gas characteristics.
For those central collisions, two dominating light charged particle sources were identified when analysing the forward part of the centre of mass (c.m.): the Projectile like (PL) source and an intermediate velocity source located at the centre of mass (c.m.) velocity. Events corresponding to a gas of free nucleons in equilibrium with clusters have been selected applying a 60$^{o}$~-~90$^{o}$ c.m. polar angular selection, so as to reduce the PL contribution, and a Coulomb corrected particle velocity selection (V$_{surf}$ larger than 3 cm/ns)~\cite{BougaultJPG19}. Fast charged particles emitted at c.m. mid-rapidity were therefore selected.
\subsection{Thermodynamical landscape}
The key observable is the velocity (V$_{surf}$) of the particles in the intermediate velocity source frame prior to
acceleration by the Coulomb field of the remaining charge. V$_{surf}$ values may be used to select different time steps of the gas evolution, fastest particles corresponding to earliest emission times. Therefore for each time step the characteristics of the evolving gas source could be deduced from the observed isotopic composition in the V$_{surf}$ bin.
\par
For each reaction, $^{136,124}$Xe+$^{124,112}$Sn, the selected data form a set of events which corresponds to an expanding gas source composed of nucleons and light clusters ($^2$H, $^3$H,  $^3$He, $^4$He and $^6$He). The associated particle V$_{surf}$ spectra are divided into 18 bins (bin width of 0.2 cm/ns). For each bin, it is possible to extract the evolving gas source characteristics and so to define 18 statistical ensembles.
The number of subsets (18) is a compromise between the width and the population of each bin. 
As described in \cite{PaisJPG47}, for each V$_{surf}$ bin the number of particles $\mathcal{N}$(A,Z) is measured except for free neutrons whose number is deduced from the proton, helion and triton numbers \cite{AlbergoNCA89}. 
The temperature (T) is estimated through the isobaric
double isotope ratio Albergo formula (T$_{HHe}$) \cite{AlbergoNCA89}. 
It has been shown in \cite{PaisJPG47} that the T$_{HHe}$ thermometer is reliable when compared to theoretical temperature values from RMF calculations. 
The mass of the gas source at the beginning of the expansion process is evaluated from a fitting procedure of the energy spectra \cite{BougaultJPG19}. This allows a direct determination of the total source mass (A$_T$) as a function of V$_{surf}$ and hence of the emission time. 
The fitting procedure showed that particle production measured between V$_{surf}$ values 3 and 4 cm/ns does not accurately represent the expanding gas because it suffers from pollution by another production source. This does not affect A$_T$ evaluation.
The baryonic density ($\rho_B$) is deduced from the total source mass and the estimation of the source volume \cite{PaisJPG47}, at the different times of the expansion. 
\par
\begin{figure}[!htbp]
\centering
  \begin{tabular}{cc}
\includegraphics[width=0.75\textwidth]{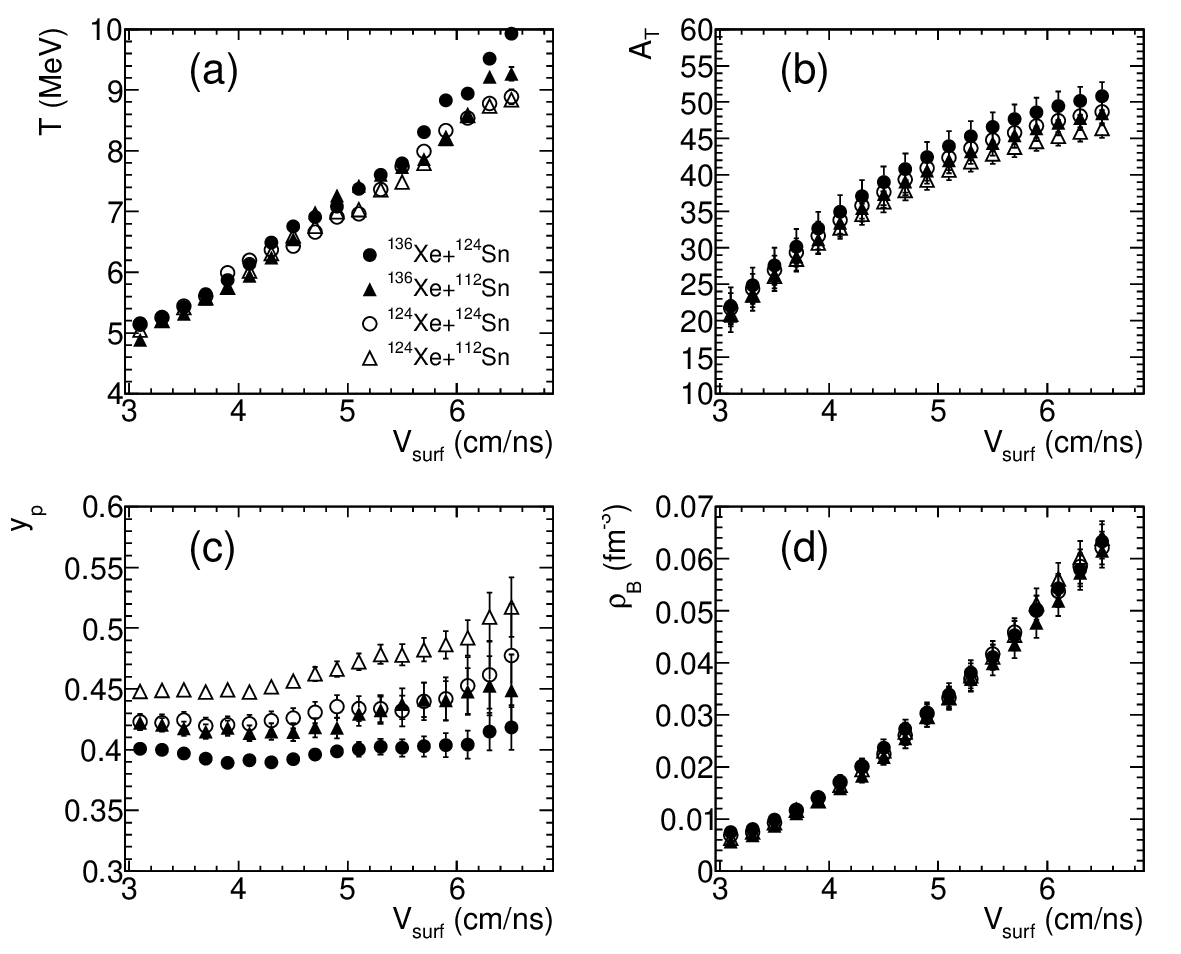} &
  \end{tabular}
\caption{Expanding gas source temperature, total mass, total proton fraction and baryon density versus V$_{surf}$ for the four reactions. Each symbol corresponds to a reaction as defined in part (a). The meaning of the symbols also applies to parts (b), (c) and (d). Each V$_{surf}$ bin corresponds to a snapshot of the time evolution of the expanding gas created in each reaction. This figure shows the mean thermodynamical characteristics of 72 statistical ensembles (18 V$_{surf}$ bins and 4 reactions).} 
\label{fig1}
\end{figure} 
\par
Figure \ref{fig1} (a-d) present respectively temperature, total mass (free neutrons, free protons and clusters), total proton fraction (total number of protons/total number of nucleons) and baryon density of the expanding gas as a function of V$_{surf}$ for the four Xe+Sn reactions. The data analysis to extract the baryon density is described in \cite{PaisPRL125, PaisJPG47}.
The temperature (T) increases with V$_{surf}$. T does not depend on the reaction (since the beam energy is the same) except for large V$_{surf}$ values, where differences of about 1 MeV between $^{136}$Xe+$^{124}$Sn and $^{124}$Xe+$^{112}$Sn reactions are observed. 
The total mass of the source increases with V$_{surf}$ and depends slightly on the reaction, the heavier the total system the larger A$_{T}$ value is.
The chemical composition of the evolving source is influenced by the total proton fraction (y$_p$). The total proton fraction values depend on the projectile plus target proton fractions leading to almost identical values for $^{136}$Xe+$^{112}$Sn and $^{124}$Xe+$^{124}$Sn reactions. Also, the hierarchy in proton fraction, based on the initial proton fraction in the input channel, is respected.
The baryonic density does not depend on the studied reaction, quasi identical values are found for each V$_{surf}$ bin.
As a function of time (i.e. decreasing V$_{surf}$) temperature, total mass and baryon density variations are compatible with an expanding gas evolution.
We have therefore 72 statistical ensembles (18 V$_{surf}$ bins and 4 reactions).
\subsection{Comment on density value extraction}
In the original work of Qin \textit{et al.} \cite{QinPRL108} and in the INDRA first analysis \cite{BougaultJPG19}, the baryon density $\rho_B$ is deduced from the experimental multiplicities using analytical expressions that explicitly assume that the physical system under study can be modeled as an ideal gas of clusters.
The density values presented in \cite{BougaultJPG19} were produced following the original method described in \cite{QinPRL108}. 
But the use of ideal gas expressions contradicts the findings that indicate the presence of in-medium effects leading to a reduction of the binding energies. In addition, this method leads to the surprising result that there is a different density value per cluster for each thermodynamical situation (for each V$_{surf}$ bin). Therefore,
modified ideal gas expressions relating the thermodynamic parameters to cluster yields have been employed for INDRA data in a subsequent analysis \cite{PaisPRL125,PaisJPG47}. The correction, estimated using Bayesian techniques, is based on the existence of a unique density regardless of the type of cluster for each thermodynamic situation.
The temperature is also obtained with the analytical hypothesis of an ideal gas but using a formula that contains a difference in binding energy which explains why this evaluation remains robust in the context of in-medium effects.
\par
For each V$_{surf}$ bin the density values are lower using the Qin \textit{et al.} \cite{QinPRL108} prescriptions but the entrance channel proton fraction independence remains. In the rest of the article, all figures will be presented as a function of V$_{surf}$ and will not, therefore, depend on the analysis method.
\section{Isoscaling framework}
This section begins by presenting, for the four studied systems, the basic observable for the isoscaling study: the chemical composition of the gas. To this end, mass fractions will be presented. The mass fraction of a given isotope is the mass of the particles corresponding to this isotope out of the total mass.
Subsequently the nuclear isoscaling formulae are presented starting from the chemical composition of a gas at equilibrium. Then some particle production scaling properties are deduced.
\subsection{Mass fractions}
\begin{figure}[!htbp]
  \centering
  \begin{tabular}{cccc} 
  \includegraphics[width=0.75\textwidth]{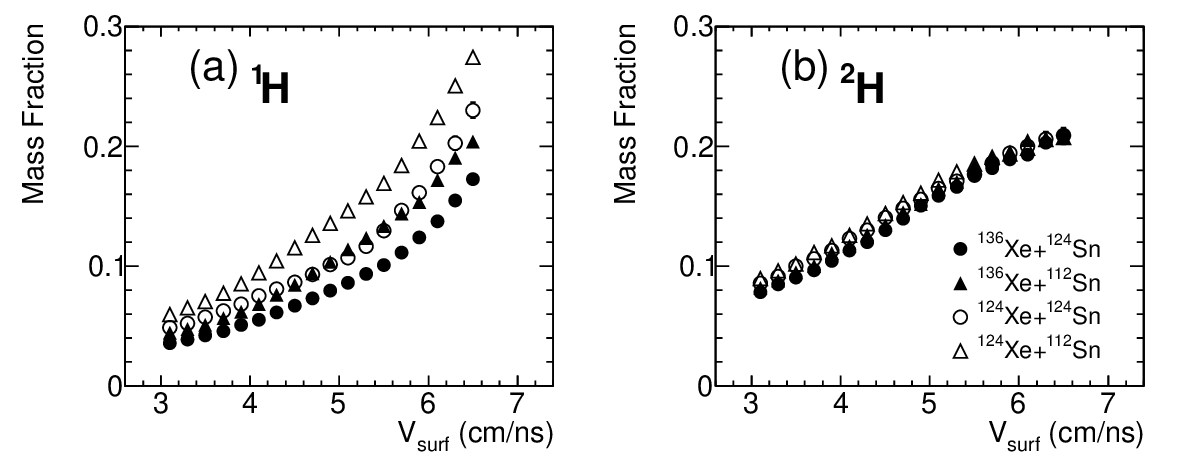} \\ 
  \includegraphics[width=0.75\textwidth]{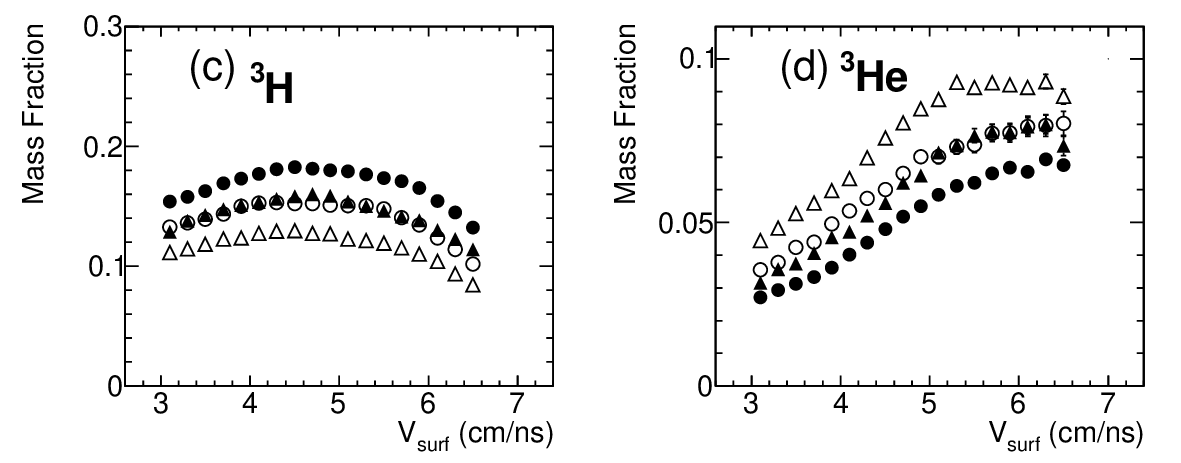} \\
  \includegraphics[width=0.75\textwidth]{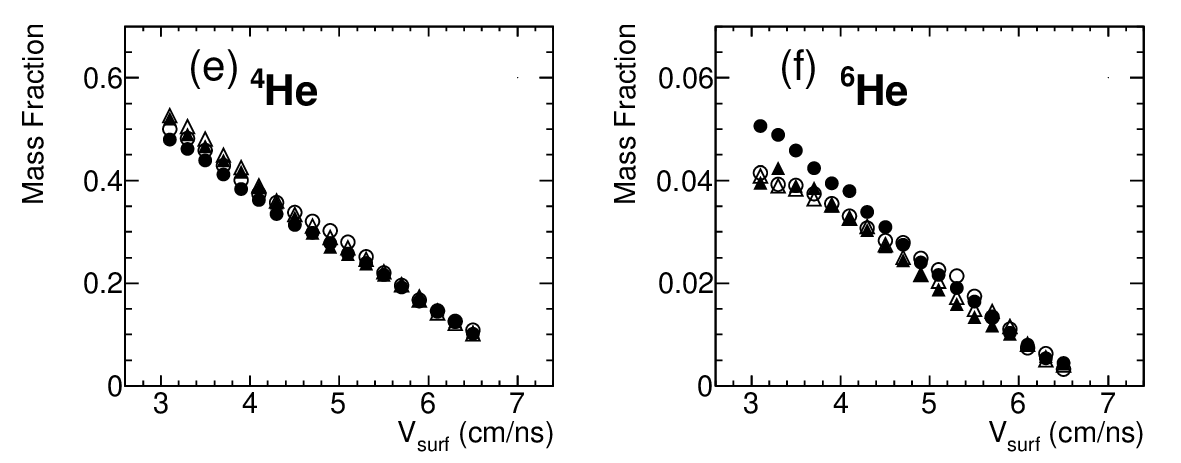}
  \end{tabular}
\caption{Mass fractions of detected charged particles as a function of V$_{surf}$ for the four reactions. Each symbol corresponds to a reaction as defined in part (b). The meaning of the symbols also applies to parts (a), (c), (d), (e) and (f). $^{1}$H mass fractions are presented in (a), $^{2}$H mass fractions in (b), $^{3}$H mass fractions in (c), $^{3}$He mass fractions in (d), $^{4}$He mass fractions in (e) and $^{6}$He mass fractions in (f). Statistical errors are in general smaller than the marker size.} 
\label{fig2}
\end{figure}

The mass fractions of detected charged particles as a function of V$_{surf}$ are shown in Figure \ref{fig2}. These are the basic observables for the following isoscaling study.
\par
\par
$^{1}$H, $^{3}$H and $^{3}$He populations depend on the neutron to proton ratio of the total projectile+target system (two different entrance channels have the same total ratio), whereas $^{2}$H, $^{4}$He populations are much less dependent on this same ratio. For $^{6}$He, the population at low V$_{surf}$ shows a difference between the more neutron-rich system and the others, while for higher V$_{surf}$ values there is hardly any difference between the four studied reactions. This $^{6}$He population evolution as a function of the N/Z ratio of the total system is not in line with other observations \cite{BougaultPRC97,Fable107} and will be discussed later.
At low V$_{surf}$ (density and temperature), the mass fraction of $^{4}$He particles is more than 50$\%$ while protons mass fraction is around 5$\%$. 
The situation is inverted for high V$_{surf}$ values since the mass fraction of $^{4}$He particles is about 10$\%$ while protons mass fraction is around 20$\%$.
As V$_{surf}$ (density and temperature) increases the $^{4}$He and $^{6}$He mass fractions decrease dramatically. $^{1}$H and $^{2}$H mass fractions are always increasing with V$_{surf}$. $^{3}$H and $^{3}$He mass fractions increase with V$_{surf}$ up to V$_{surf}\approx$5 cm/ns.
\par
The thermodynamic landscape and the isotopic composition of the different statistical ensembles is now established.
\subsection{Isoscaling ratio}
In a chemical equilibrium between free nucleons and clusters within an interaction
volume $V$ and imposing a thermal equilibrium at a temperature T with a Maxwell-Boltzmann momentum distribution, the number of particle (A,Z) of such a system is given by \cite{PaisJPG47}:
\begin{equation}
\mathcal{N}(A,Z)~=~V~\frac {[M(A,Z)~2\pi~T]^{3/2}~g(A,Z)}{h^{3}}~\exp \left [ \frac {\mu(A,Z)}{T} \right ],
\label{eq1}
\end{equation}
where \textit{M(A,Z)} is the mass of particle (A,Z) and \textit{h} is the Planck constant.
The internal partition sum, neglecting the population of excited states, reads:
\begin{equation}
g(A,Z)~=~[2J(A,Z)+1]~\exp\left [ \frac{B(A,Z)}{T}\right ].
\label{eq2}
\end{equation}
$B(A,Z)$ is the binding energy of the ground state of particle \textit{(A,Z)} which, in case of in-medium effects, depends on temperature and density and \textit{J(A,Z)} is the spin. $M(A,Z)~=~Am~-~B(A,Z)$, with m the nucleon mass, depends also on temperature and density in case of in-medium effects. 
\par
The chemical potential of particle \textit{(A,Z)}, $\mu(A,Z)$, is related to the number of proton \textit{(Z)} and the number of neutrons \textit{(N=A-Z)} through the following relationship involving the neutron and the proton chemical potentials:
\begin{equation}
\mu(A,Z)~=~N~\mu(1,0)~+~Z~\mu(1,1),
\label{eq3}
\end{equation}
where $\mu(A,Z)$ value depends on the thermodynamical conditions: temperature \textit{(T)}, density ($\rho$) and total proton fraction of the system ($y_p$).
\par
Taking two statistical ensembles characterized by (\textit{T$_{S1}$}, $\rho_{S1}$, $y_{pS1}$) and (\textit{T$_{S2}$}, $\rho_{S2}$, $y_{pS2}$), the ratio of the number of particles \textit{(A,Z)} between situations S2 and S1 is given by,
\begin{eqnarray}
\fl
R(A,Z)_{S2S1}~\equiv~\frac {\mathcal{N}(A,Z)_{S2}} {\mathcal{N}(A,Z)_{S1}}~=
~\left ( \frac {V_{S2}}{V_{S1}}\right )~\left ( \frac{T_{S2}}{T_{S1}}\right )^{3/2} 
~\left ( \frac {Am~-~B(A,Z)_{S2}}{Am~-~B(A,Z)_{S1}}\right )^{3/2} \nonumber\\
\exp\left [ \left(\frac {\mu(1,0)_{S2}}{T_{S2}}~-~\frac {\mu(1,0)_{S1}}{T_{S1}}\right)~N
~+~\left(\frac {\mu(1,1)_{S2}}{T_{S2}}~-~\frac {\mu(1,1)_{S1}}{T_{S1}}\right)~Z \right ] \nonumber\\
\exp\left [ \frac{B(A,Z)_{S2}}{T_{S2}}~-~\frac{B(A,Z)_{S1}}{T_{S1}}\right ].
\label{eq4}
\end{eqnarray}
\par
In case where situations S1 and S2 differ only in the total proton fraction, we have $T_{S1}=T_{S2}=T$, $V_{S1}=V_{S2}$ and 
$B(A,Z)_{S1}=B(A,Z)_{S2}$. The ratio of the number of particles \textit{(A,Z)} between situations S2 and S1 is then given by,
\begin{equation}
R(A,Z)_{S2S1}
~=~
\exp\left [ \alpha~N+\beta~Z \right ],
\label{eq5}
\end{equation}
$\alpha$ and $\beta$ being variables which are independent of \textit{N} and \textit{Z}.
In that case, this scaling law which relates ratios of isotope
yields obeys an exponential dependence
on the neutron and proton numbers of the isotopes
with $\alpha=[\mu(1,0)_{S2}-\mu(1,0)_{S1}]/T=\Delta \mu(1,0)/T$ and $\beta=[\mu(1,1)_{S2}-\mu(1,1)_{S1}]/T=\Delta \mu(1,1)/T$. This scaling law is known as Isoscaling and the ratio is noted \textit{R(A,Z)}$_{S2S1}$. It should be noted that this relationship holds in the case of in-medium effects.
\par
In the case where the two situations S1 and S2 differ slightly in total size, the condition $V_{S1}=V_{S2}$ may not be fulfilled and in this case, equation (\ref{eq5}) becomes equation (\ref{eq6}) with a normalisation constant \textit{C} that is also independent of \textit{N} and \textit{Z}. 
\begin{equation}
R(A,Z)_{S2S1}
~=~
C~
\exp\left [ \alpha~N+\beta~Z \right ]
\label{eq6}
\end{equation}
\par
From a theoretical point of view, a situation of thermal and chemical equilibrium implies the observation of isoscaling.
\subsection{Isoscaling ratio scaling properties} \label{ScalingProperties}
If equation \ref{eq5} holds, then we have the following properties between isoscaling ratios of particles \textit{(A,Z)}, \textit{(A',Z')}, \textit{(A+A',Z+Z')} and \textit{(A-A',Z-Z')} \cite{Chajecki2014}:
\begin{eqnarray}
R(A+A',Z+Z')_{S2S1}~=~R(A,Z)_{S2S1}~R(A',Z')_{S2S1} \nonumber\\
R(A-A',Z-Z')_{S2S1}~=~R(A,Z)_{S2S1}~/~R(A',Z')_{S2S1},
\label{eq7}
\end{eqnarray}
therefore, 
\begin{eqnarray}
R(4,2)_{S2S1}~=~R(2,1)_{S2S1}^{2}~=~R(3,1)_{S2S1}~R(1,1)_{S2S1} 
\nonumber
\\ R(1,1)_{S2S1}~=~\frac{R(3,2)_{S2S1}}{R(2,1)_{S2S1}}.
\label{eq8}
\end{eqnarray}
The second part of equation \ref{eq7} defines a reduced isotopic ratio which will be discussed further below. 
\par
In the case where the two situations S1 and S2 differ slightly in total size, equation \ref{eq6} is more general and therefore equations \ref{eq8} become
\begin{eqnarray}
R(4,2)_{S2S1}~=~\frac{R(2,1)_{S2S1}^{2}}{C}~=~\frac{R(3,1)_{S2S1}~R(1,1)_{S2S1}}{C} \nonumber
\\ 
R(1,1)_{S2S1}~=~C~\frac{R(3,2)_{S2S1}}{R(2,1)_{S2S1}}.
\label{eq9}
\end{eqnarray}
These are the four basic mathematical equations which connect $^{1}$H, $^{2}$H, $^{3}$H, $^{3}$He and $^{4}$He production ratios in situations of thermal and chemical equilibrium.
\section{Isoscaling results}
In this section, data is confronted with the isoscaling law and scaling properties. Then 
the last word is given using the isoscaling law with reduced isotopic ratios.
\subsection{Isoscaling ratio}
Ratios between two Xe+Sn reactions are computed for each of the 18 bins in V$_{surf}$ in order to compare as much as possible identical equilibrium situations ($\rho_{B}$, T).
\begin{figure}[!htbp]
\centering
  \begin{tabular}{cc}
\includegraphics[width=0.75\textwidth]{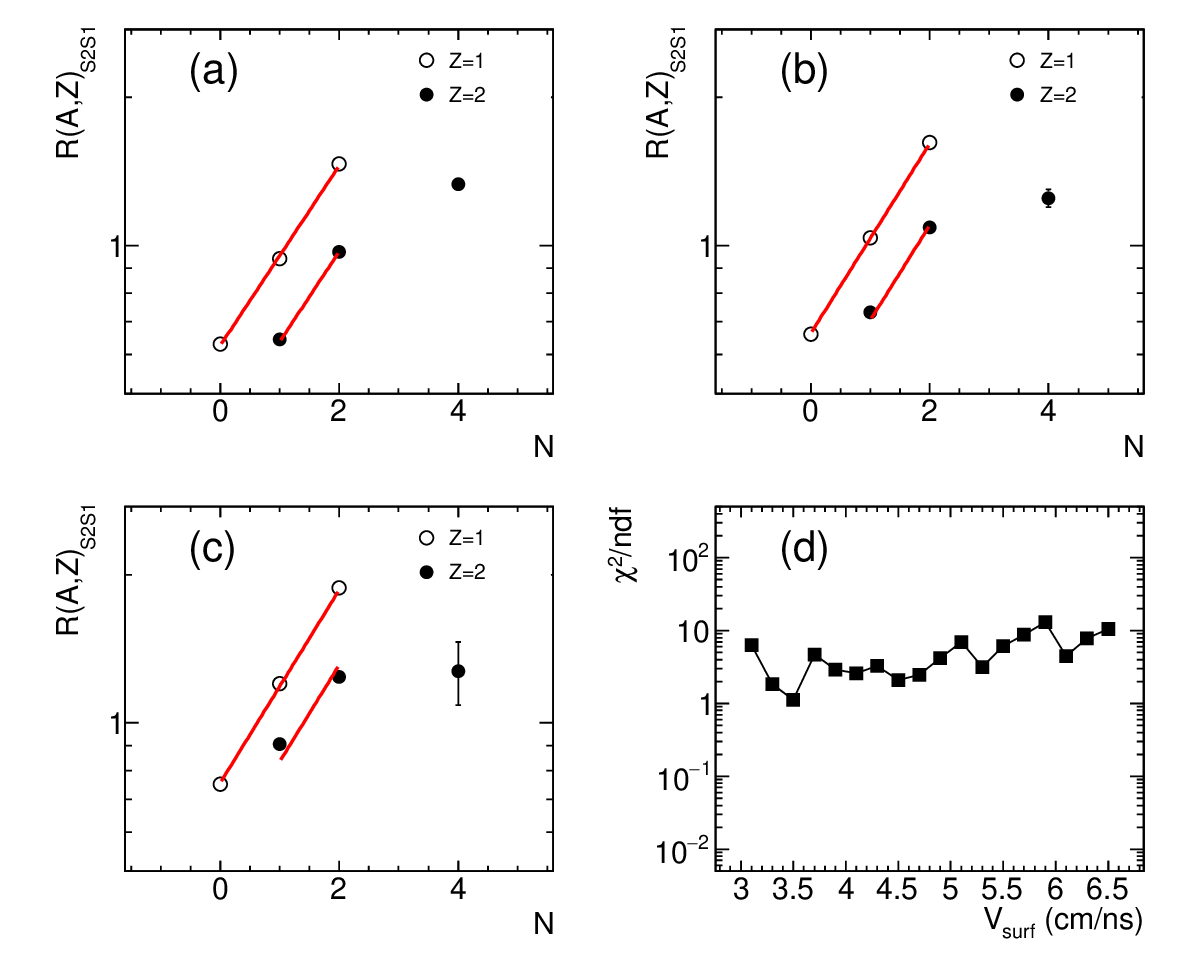} 
  \end{tabular}
\caption{Isoscaling for statistical ensembles from $^{136}$Xe+$^{124}$Sn reaction (S2) and statistical ensembles from $^{124}$Xe+$^{112}$Sn reaction (S1). (a), (b) and (c) present the ratios of the number of particles between S2 and S1 for $^{1}$H, $^{2}$H, $^{3}$H,
$^{3}$He, $^{4}$He and $^{6}$He (open circles are Hydrogen isotopes and filled circles are Helium isotopes). (a) corresponds to V$_{surf}$ bin number 2 (V$_{surf}$=3.3 $\pm$ 0.1 cm/ns), (b) corresponds to V$_{surf}$ bin number 9 (V$_{surf}$=4.7 $\pm$ 0.1 cm/ns) and (c) corresponds to V$_{surf}$ bin number 17 (V$_{surf}$=6.3 $\pm$ 0.1 cm/ns). The lines in (a), (b) and (c) are the fitting result using equation \ref{eq6} excluding $^{6}$He data. (d) presents the $\chi^{2}$ values of the 18 independent fits divided by the number of degrees of freedom (ndf).} 
\label{fig3}
\end{figure} 
Figure \ref{fig3}-a, \ref{fig3}-b and \ref{fig3}-c present $R(A,Z)_{S2S1}$ data for three V$_{surf}$ bins considering $^{136}$Xe+$^{124}$Sn reaction as equilibrium situation S2 and $^{124}$Xe+$^{112}$Sn reaction as equilibrium situation S1. These two reactions are chosen because they lead to the largest difference in total proton fraction for each bin in V$_{surf}$.
$^{6}$He ratios are absolutely not in phase with the isoscaling property of equation \ref{eq6} (this is true for all 18 bins).
This phenomenon of deviation from the isoscaling law of $^{6}$He has been observed in evaporation models \cite{TsangEur2006} but this is not a general feature (see \cite{TsangPRC64} for example). This may be due to finite size effects (lack of neutrons in the expanding gas, angular momentum conservation,...) and could explain the atypical behaviour of $^{6}$He observed in figure \ref{fig2}. This phenomenon could also be explained 
by the very pronounced halo structure of $^{6}$He, coupled with the very weak binding energy of the two extra 
neutrons forming the halo \cite{LagoyannisPLB518, SunPLB518}. 
The consequence is that the equilibrium constant measurements for $^6$He published in \cite{PaisPRL125,PaisJPG47,BougaultJPG19} must be taken with caution with respect to infinite nuclear matter.
\par
Equation \ref{eq6} is used in order to fit particle production ratios between the two Xe+Sn reactions per bin of V$_{surf}$. 
The adopted procedure is the following: for each V$_{surf}$ bin, the fit (equation \ref{eq6}) is performed 
excluding $^{6}$He, therefore with five points.
Figure \ref{fig3}-d shows the $\chi^{2}$/ndf extracted from the fits for the 18 bins in V$_{surf}$. Fits are presented as lines in figure \ref{fig3}-a, b and c for the three selected V$_{surf}$ bins. 
The isoscaling property is observed but with $\chi^{2}$/ndf values that reflect some deviations (\textit{Z}=2 in Figure \ref{fig3}-c for example). We conclude here that the quality of the fit is far from perfect for most bins. To take into account total mass difference between situations S2 and S1 we used equation \ref{eq6} instead of equation \ref{eq5} but it appears that this is not enough for some bins. The same conclusions are drawn by examining the other combinations of reactions.
In the following we will concentrate on $^{136}$Xe+$^{124}$Sn and $^{124}$Xe+$^{112}$Sn reactions. 
%
\subsection{Isoscaling ratio scaling properties}
The question of the isoscaling law can be examined by using scaling properties of number of particle ($\mathcal{N}$(A,Z)) ratios presented in the subsection \ref{ScalingProperties}. 
\begin{figure}[!htbp]
\centering
  \begin{tabular}{cc}
\includegraphics[width=0.75\textwidth]{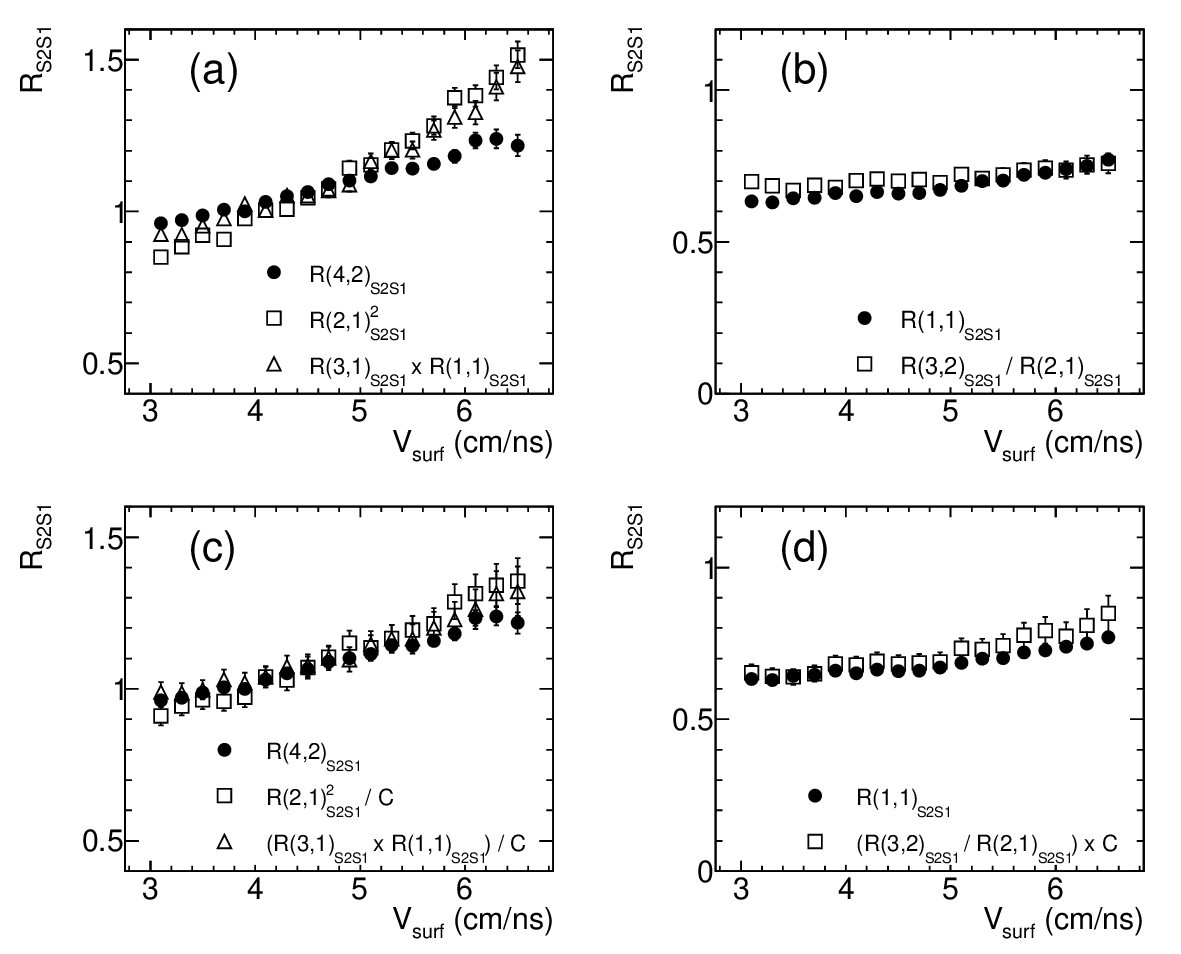} 
  \end{tabular}
\caption{Scaling properties from equations \ref{eq8} are presented on top (a, b). Scaling properties from equations \ref{eq9} are presented on bottom (c, d) 
using the normalisation constant values of isoscaling fits.
Ratio and ratio products are presented as a function of V$_{surf}$. S2 stands for statistical ensembles from $^{136}$Xe+$^{124}$Sn reaction and S1 for statistical ensembles from $^{124}$Xe+$^{112}$Sn reaction.} 
\label{fig4}
\end{figure} 
\par
Figure \ref{fig4} shows the ratios, ratio products and ratio divisions of equations (\ref{eq8}) ((a) and (b)) and of equation (\ref{eq9}) ((c) and (d)). Ratios and ratio products linked to $^{4}$He production are presented on the left-side while ratios and ratio division linked to $^{1}$H production are presented on the right-side. If the isoscaling law is verified, the different points should be superimposed for the two types of equations. The total mass effect is visible since there exists large discrepancies in figure \ref{fig4}-a, especially for V$_{surf}$ greater than 5 cm/ns where differences of total mass are most significant. This justifies the use of equation \ref{eq6} instead of equation \ref{eq5}.
This effect is not visible for 
figure \ref{fig4}-b because $^{1}$H ratio is not compared to ratio products but to ratio division and thus the total mass effect is not amplified.
\par
We will now concentrate on equations \ref{eq9} deduced from equation \ref{eq6} (figures \ref{fig4}-c and -d).
Equalities concerning $^{4}$He ratio are much better satisfied, while the one connected to $^{1}$H ratio is now better verified for low V$_{surf}$ values. 
It was shown in \cite{BougaultJPG19} that particle production from the expanding source between V$_{surf}$=3 cm/ns and 4 cm/ns suffers from uncertainty due to additional production from another source. This effect is clearly visible in figure \ref{fig4}-c. It does not clearly appear in figure \ref{fig4}-d probably because of the use of ratio division.  
For large values of V$_{surf}$ there exists a slight discrepancy. We will show in the next section that this effect is due to a difference in temperature between S2 and S1 (see figure \ref{fig1}-a).
\par
We add some comments concerning $^{4}$He and $^{1}$H ratios (filled circles in figures \ref{fig4}-c and -d). Contrary to what figure \ref{fig2} might suggest comparing mass fractions for the two reactions, $^{4}$He ratio evolves with V$_{surf}$. This evolution is linked to temperature and density variations and also to a decrease of the total proton fraction difference between the two systems as V$_{surf}$ decreases (figure \ref{fig1}). In absolute value $^{1}$H ratio varies less than $^{4}$He ratio. But one should not conclude that $^{1}$H ratio is almost constant because in relative value the variation of the two isotope ratios is equivalent.
\subsection{Isoscaling using reduced isotopic ratios}
There exist differences in the size of the expanding gas between the different reactions. This point has been circumvented by using equation \ref{eq6} instead of equation \ref{eq5} for the time being.
To eliminate the extra particle production problem between V$_{surf}$=3 cm/ns and 4 cm/ns and with the aim of using a more genuine fit procedure, we now use the reduced isotopic ratio  
presented in the subsection \ref{ScalingProperties}.
The reduced isotopic ratio for
particle \textit{(A,Z)} is the particle ratio normalized with respect to the ratio for a given isotope. 
This technique makes it possible to eliminate data uncertainties between two statistical ensembles \cite{BotvinaPRC65,GeraciNPA732}.
The normalization isotope is chosen to be $^{1}$H in our case. 
\begin{figure}[!htbp]
\centering
  \begin{tabular}{cc}
\includegraphics[width=0.75\textwidth]{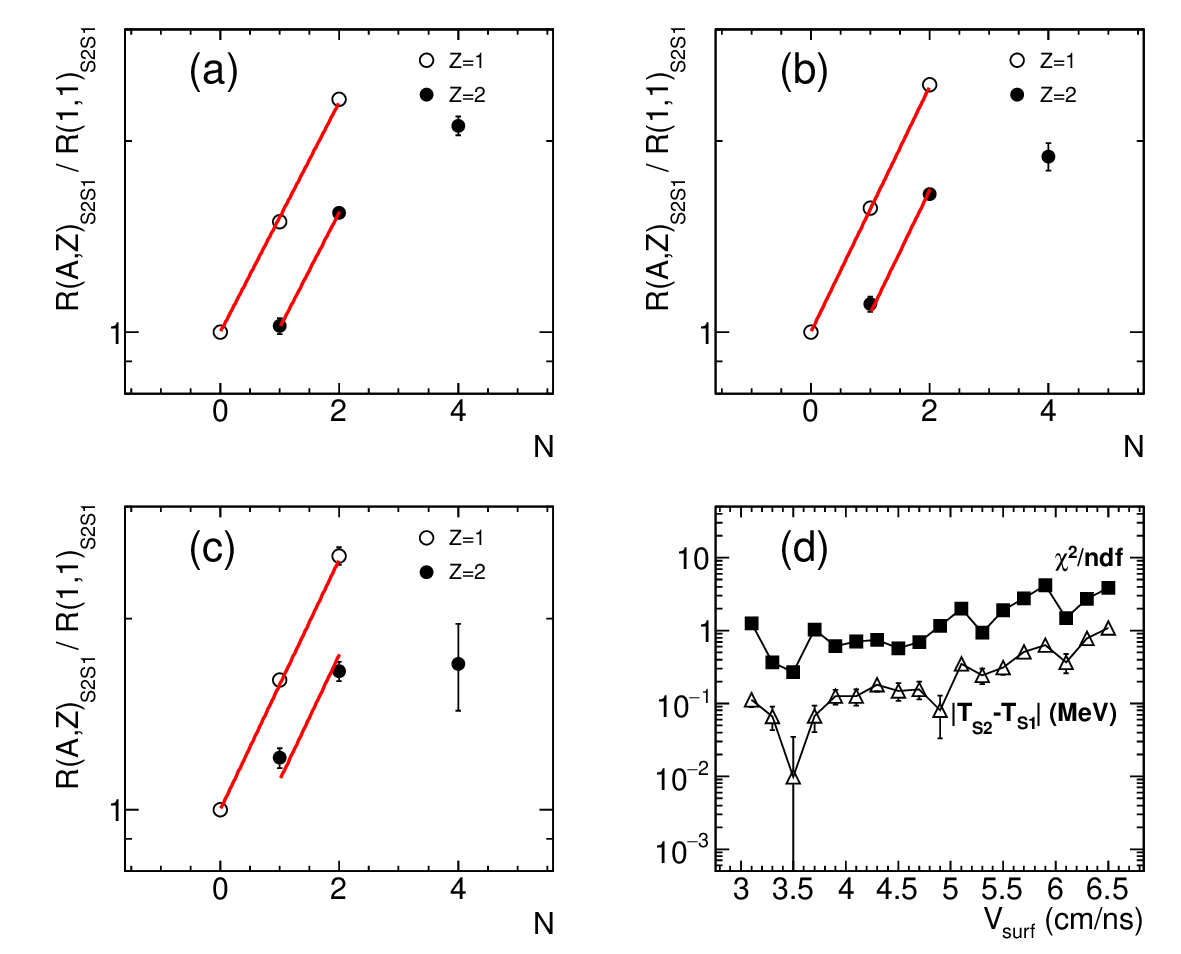}
  \end{tabular}
\caption{Double ratio isoscaling for statistical ensembles from $^{136}$Xe+$^{124}$Sn reaction (S2) and statistical ensembles from $^{124}$Xe+$^{112}$Sn reaction (S1). (a), (b) and (c) present the ratios of the number of particles between S2 and S1 for $^{1}$H, $^{2}$H, $^{3}$H,
$^{3}$He, $^{4}$He and $^{6}$He normalized to $^{1}$H ratio (open circles are Hydrogen isotopes and filled circles are Helium isotopes). (a) corresponds to V$_{surf}$ bin number 2 (V$_{surf}$=3.3 $\pm$ 0.1 cm/ns), (b) corresponds to V$_{surf}$ bin number 9 (V$_{surf}$=4.7 $\pm$ 0.1 cm/ns) and (c) corresponds to V$_{surf}$ bin number 17 (V$_{surf}$=6.3 $\pm$ 0.1 cm/ns). The lines in (a), (b) and (c) are the fitting result using equation \ref{eq10} excluding $^{6}$He data. (d) presents the $\chi^{2}$/ndf values of the 18 independent fits (filled squares) and the absolute value of the temperature difference between S2 and S1 (open triangles).} 
\label{fig5}
\end{figure} 
\par
The reduced isotopic ratio leads to, 
\begin{equation}
\frac {R(A,Z)_{S2S1}} {R(1,1)_{S2S1}}
~=~
\exp\left [ \alpha~N+\beta~(Z-1) \right ]
\label{eq10}
\end{equation}
The reduced isotopic ratio for protons is equal to 1 by construction and
the normalisation constant is eliminated. The fits are thus made with only two parameters.
\par
Reduced isotopic ratios for Hydrogen and Helium isotopes are presented in figures \ref{fig5}-a, \ref{fig5}-b and \ref{fig5}-c. As before, $^{136}$Xe+$^{124}$Sn and $^{124}$Xe+$^{112}$Sn reactions are chosen. The layout is similar to that in figure \ref{fig3}: points represent the data and lines are the result of the fit (equation \ref{eq10}) for three bins in V$_{surf}$. The production of $^{6}$He is still problematic, and the use of double ratios is not intended to fix this problem.
In figure \ref{fig5}-d are presented the $\chi^{2}$/ndf and the difference in temperature between the two systems as a function of V$_{surf}$. First, 
the quality of the fits (as indicated by the $\chi^{2}$/ndf) is significantly improved compared to before (Figure \ref{fig3}), hence the need to use reduced isotopic ratios especially for low V$_{surf}$ values. Secondly, we observe that, for $\chi^{2}$/ndf and the temperature difference, global behaviours are similar: the $\chi^{2}$/ndf increases as the temperature difference increases and vice-versa. This is consistent because the equation \ref{eq10} is valid only under the condition $T_{S2}$=$T_{S1}$. When the temperature difference reaches values approaching one MeV, $\chi^{2}$/ndf value increases. This relationship between $\chi^{2}$/ndf and temperature difference 
shows that the data are compatible with the isoscaling law as long as the temperatures of the two systems are close. 
\par
We conclude that the data are compatible with the equilibrium hypothesis. The lesser compatibility of the data at large V$_{surf}$ with the isoscaling law is rather well explained by the differences in temperature between the studied systems.
\par
\begin{figure}[!htbp]
\centering
  \begin{tabular}{cc}
\includegraphics[width=0.75\textwidth]{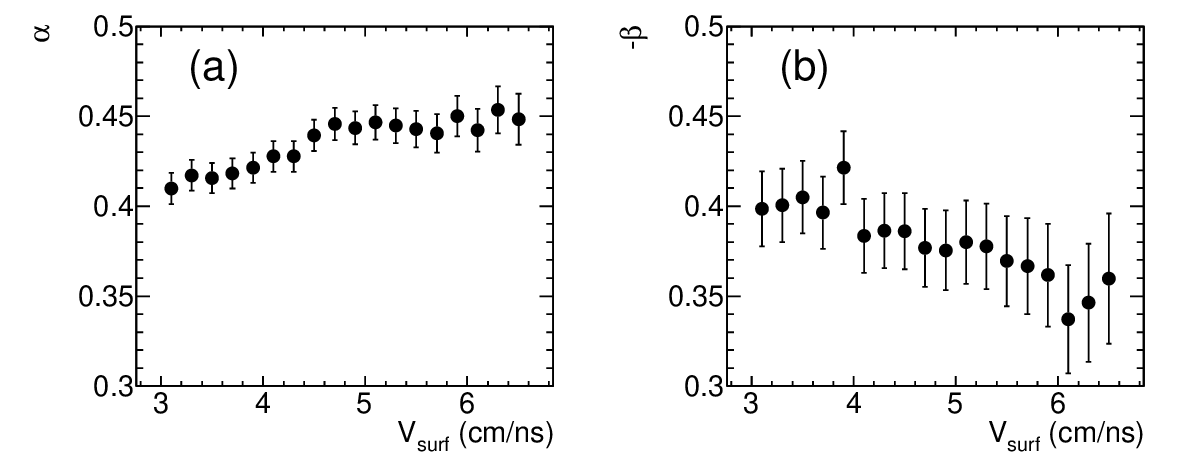}
  \end{tabular}
\caption{Fit parameter values (equation \ref{eq10} and figure \ref{fig5}) for the 18 bins in V$_{surf}$. $\alpha$ is related to neutron number and $\beta$ to proton number of the different isotopes. $\beta$ is plotted as -$\beta$ for easier comparison with alpha values. Statistical ensembles from $^{136}$Xe+$^{124}$Sn and $^{124}$Xe+$^{112}$Sn reactions are used.}
\label{fig6}
\end{figure}

Finally, we present in figure \ref{fig6} the values of the isoscaling parameters $\alpha$ and $\beta$ extracted from the fits.
Going back to figure \ref{fig5}, $\alpha$ is the slope of the affine functions in the semi logarithmic scale representation and $\beta$ is the distance between elements.
We can see that the value of the parameter $\alpha$ increases with V$_{surf}$ until it reaches a limit value at V$_{surf}=5~$cm/ns. This saturation effect should be treated with caution, as the temperatures of the two systems differ from this value of V$_{surf}$ onwards. Referring to publication \cite{TsangPRL86}, we note that the values are compatible with those found experimentally for low-density de-excitation mechanisms (multifragmentation). 
The $\alpha$ parameter values are also compatible with theoretical multifragmentation calculations carried out for systems with similar proton fraction values (figure 9 of \cite{TsangEur2006}). 
The $\beta$ parameter values are subject to a larger error bar due to the limited number of isotopes used for their determination. The values are close to those of the $\alpha$ parameter but are not equal in absolute value. The sum of $\alpha$ and $\beta$ tends to increase with $V_{surf}$.
A detailed study of those parameters
and a comparison with an isoscaling analysis using the RMF model will be the subject of a following publication.
\section{Conclusion}
We have presented a quantitative study of the properties and consequences of the isoscaling law with experimental data. These data concern hot nuclear systems at low density and finite temperature.
In our article isoscaling is assessed quantitatively, not by inspecting figures visually. Here $\chi^{2}$ values are given and the deviations are studied step by step.
The deviations concern different effects: the difference in total mass and temperature of the studied systems as well as uncertainty effects. After studying isoscaling through the usual formula, we concluded that it is better to use the reduced isotopic ratio which allows to erase some differences (total mass and uncertainty). This ratio does not eliminate all effects. However, the extracted $\chi^{2}$ values are closely correlated to the temperature difference values. For temperature differences between the studied systems of around 1 MeV, the isoscaling fit does not reproduce the data well.  This gives a confidence interval for future studies related to isoscaling. The deviations observed with regard to the isoscaling law are in the same direction as the studies published in \cite{SouzaPRC80, SouzaPRC106}.
\par
This work is a complementary study to our previous publications on hot and low density nuclear systems \cite{PaisPRL125, PaisJPG47,BougaultJPG19}.
It provides further confirmation of the compatibility of the data with the thermal and chemical equilibrium hypothesis and thus confirms the validity of using thermal model framework.
\par
However, the analysis indicates that extrapolation of detected $^{6}$He production to infinite nuclear matter should be taken with caution.
The observed discrepancies concerning isoscaling for the $^{6}$He could be explained by the difficulty to form such a neutron-rich nucleus within a small size system (between 20 and 50 nucleons) or by the specific halo nature of this cluster. This point needs to be verified with reactions that allow access to gaseous nuclear systems containing higher mass clusters. 


\section*{References}
\begin{thebibliography}{50} 
\bibitem{TsangPRC64bis} Tsang M.B., Gelbke C.K., Liu X.D., Lynch W.G., Tan W.P., Verde G., Xu H.S., Friedman W.A., Donangelo R., Souza S.R., Das C.B., Das Gupta S., and Zhabinsky D. 2001 Phys. Rev. C {\bf 64} 054615

\bibitem{XuPRL85} Xu H.S., Tsang M.B., Liu T.X., Liu X.D., Lynch W.G., Tan W.P., Vander Molen A., Verde G., Wagner A., Xi H.F., Gelbke C.K., Beaulieu L., Davin B., Larochelle Y., Lefort T., de Souza R.T., Yanez R., Viola V.E., Charity R.J. and Sobotka L.G. 2000 Phys. Rev. Lett. {\bf85} 716

\bibitem{TsangPRL86} Tsang M.B., Friedman W.A., Gelbke C.K., Lynch W.G., Verde G., and Xu H.S. 2001 Phys. Rev. Lett. {\bf86} 5023

\bibitem{TsangEur2006} Tsang M.B.,Bougault R., Charity R., Durand D., Friedman W.A., Gulminelli F., Le F{\'e}vre A., Raduta Al.H., Raduta Ad.R., Souza S., Trautmann W., and Wada R. 2006 Eur. Phys. J. A {\bf30} 129

\bibitem{TsangPRC64} Tsang M.B., Friedman W.A., Gelbke C.K., Lynch W.G., Verde G., and Xu H.S. 2001 Phys. Rev. C {\bf 64} 041603(R)

\bibitem{BotvinaPRC65} Botvina A.S., Lozhkin O.V., and Trautmann W. 2002 Phys. Rev. C {\bf 65} 044610

\bibitem{OnoPRC68} Ono A., Danielewicz P., Friedman W.A., Lynch W.G., and Tsang M.B. 2003 Phys. Rev. C {\bf 68} 051601(R)

\bibitem{SouzaPRC80} Souza S.R., Tsang M.B., Carlson B.V., Donangelo R., Lynch W.G., and A. W. Steiner A.W. 2009 Phys. Rev. C {\bf 80} 044606

\bibitem {PaisPRL125} Pais H., Bougault R., Gulminelli F., Provid{\^e}ncia C. , Bonnet E., Borderie B., Chbihi A., Frankland J.D., Galichet E., Gruyer D., Henri M., Le Neindre N., Lopez O., Manduci L., P{\^a}rlog M., and Verde G. 2020 Phys. Rev. Lett. {\bf 125} 012701 

\bibitem {PaisJPG47} Pais H., Bougault R., Gulminelli F., Provid{\^e}ncia C. , Bonnet E., Borderie B., Chbihi A., Frankland J.D., Galichet E., Gruyer D., Henri M., Le Neindre N., Lopez O., Manduci L., P{\^a}rlog M., and Verde G. 2020 Journ. Phys. G {\bf47} 105204

\bibitem{PaisPRC97} Pais H., Gulminelli F., Provid{\^e}ncia C., and R{\"o}pke G. 2018 Phys. Rev. C {\bf 97} 045805

\bibitem{Pouthas1995} Pouthas J., Borderie B., Dayras R., Plagnol E., Rivet M.F., Saint-Laurent F., Steckmeyer J.C., Auger G., Bacri C.O., Barbey S., Barbier A., Benkirane A.,
Benlliure J., Berthier B., Bougamont E., Bourgault P., Box P., Bzyl R.,
Cahan B., Cassagnou Y., Charlet D., Charvet J.L., Chbihi A., Clerc T.,
Copinet N., Cussol D., Engrand M., Gautier J.M., Huguet Y., Jouniaux O.,
Laville J.L., Le Botlan P., Leconte A., Legrain R., Lelong P., Le Guay M.,
Martina L., Mazur C., Mosrin P., Olivier L., Passerieux J.P., Pierre S., Piquet B.,
Plaige E., Pollacco E.C., Raine B., Richard A., Ropert J., Spitaels C., Stab L.,
Sznajderman D., Tassan-got L., Tillier J., Tripon M., Vallerand P., Volant C.,
Volkov P., Wieleczko J.P., and Wittwer G. 1995 Nucl. Instr. and Meth. A {\bf 357} 418

\bibitem{BougaultPRC97} Bougault R., Bonnet E., Borderie B., Chbihi A., Dell{'}Aquila D., Fable Q., Francalanza L., Frankland J.D., Galichet E., Gruyer D., Guinet D., Henri M., La Commara M., Le Neindre N., Lombardo I., Lopez O., Manduci L., Marini P., P{\^a}rlog M., Roy R., Saint-Onge P., Verde G., Vient E., and Vigilante M. 2018 Phys. Rev. C {\bf 97} 024612

\bibitem{BougaultJPG19} Bougault R., Bonnet E., Borderie B., Chbihi A., Frankland J.D., Galichet E., Gruyer D., Henri M., La Commara M., Le Neindre N., Lombardo I., Lopez O.,
Manduci L., P{\^a}rlog M., Roy R., Verde G., and Vigilante M. 2020 Journ. Phys. G {\bf47} 025103

\bibitem{AlbergoNCA89} Albergo S., Costa S., Costanzo E., and Rubbino A. 1985 Nuovo Cimento A {\bf89} 1

\bibitem{QinPRL108} Qin L., Hagel K., Wada R., Natowitz J.B., Shlomo S., Bonasera A., R{\"o}pke G., Typel S., Chen Z., Huang M., Wang J., Zheng H., Kowalski S., Barbui M., Rodrigues M.R.D., Schmidt K., Fabris D., Lunardon M., Moretto S., Nebbia G., Pesente S., Rizzi V., Viesti G., Cinausero M., Prete G., Keutgen T., El Masri Y., Majka Z., and Ma Y.G. 2012 Phys. Rev. Lett. {\bf 108} 172701

\bibitem{Fable107} Fable Q., Chbihi A., Frankland J.D., Napolitani P., Verde G., Bonnet E., Borderie B., Bougault R., Galichet E., G{\'e}nard T., Gruyer D., Henri M., La Commara M., Le F{\'e}vre A., Lemari{\'e} J., Le Neindre N., Lopez O., Marini P., P{\^a}rlog M., Rebillard-Souli{\'e} A., Trautmann W., Vient E., and Vigilante M. 2023 Phys. Rev. C {\bf107} 014604

\bibitem{Chajecki2014} Chajecki Z., Youngs M., Coupland D.D.S., Lynch W.G., Tsang M.B., Brown D., Chbihi A., Danielewicz P., deSouza R.T., Famiano M.A., Ghosh T.K., Giacherio B., Henzl V., Henzlova D., Herlitzius C., Hudan S., Kilburn M.A., Lee Jenny, Lu F., Lukyanov S., Rogers A.M., Russotto P., Sanetullaev A., Showalter R.H., Sobotka L.G., Sun Z.Y., Vander Molen A.M., Verde G., Wallace M.S., and Winkelbauer J. 2014 arXiv:1402.5216 [nucl-ex]

\bibitem{LagoyannisPLB518} Lagoyannis A., Auger F., Musumarra A., Alamanos N., Pollacco E.C., Pakou A., Blumenfeld Y., Braga F., La Commara M., Drouart A., Fioni G., Gillibert A., Khan E., 
Lapoux V., Mittig W., Ottini-Hustache S., Pierroutsakou D., Romoli M., Roussel-Chomaz P.,
Sandoli M., Santonocito D., Scarpaci J.A., Sida J.L., and Suomij{\"a}rvi T. 2001 Phys. Lett. B {\bf 518} 27

\bibitem{SunPLB518} Sun Y.L., Nakamura T., Kondo Y., Satou Y., Lee J., Matsumoto T., Ogata K.,
Kikuchi Y., Aoi N., Ichikawa Y., Ieki K., Ishihara M., Kobayshi T.,
Motobayashi T., Otsu H., Sakurai H., Shimamura T., Shimoura S., 
Shinohara T., Sugimoto T., Takeuchi S., Togano Y., Yoneda K. 2021 Phys. Lett. B {\bf 814} 136072

\bibitem {GeraciNPA732} Geraci E., Bruno M., D{'}Agostino M., De Filippo E.,
Pagano A., Vannini G., Alderighi M., Anzalone A.,
Auditore L., Baran V.,f, Barn{\'a} R., Bartolucci M., Berceanu I.,
Blicharska J., Bonasera A., Borderie B., Bougault R.,
Brzychczyk J., Cardella G., Cavallaro S., Chbihi A., Cibor J.,
Colonna M., De Pasquale D., Di Toro M., Giustolisi F.,
Grzeszczuk A., Guazzoni P., Guinet D., Iacono-Manno M.,
Italiano A., Kowalski S., La Guidara E., Lanzalone G.,
Lanzan{\`o} G., Le Neindre N., Li S., Lo Nigro S., Maiolino C.,
Majka Z., Manfredi G., Paduszynski T., Papa M., Petrovici M.,
Piasecki E., Pirrone S., Politi G., Pop A., Porto F., Rivet M.F.,
Rosato E., Russo S., Russotto P., Sechi G., Simion V.,
Sperduto M.L., Steckmeyer J.C., Trifir{\'o} A., Trimarchi M.,
Vigilante M., Wieleczko J.P., Wilczynski J., Wuo H., Xiaoo Z.,
Zetta L, and Zipper W. 2004 Nuclear Physics A {\bf 732} 173

\bibitem{SouzaPRC106} Souza S.R., Donangelo R., Lynch W.G., and Tsang M.B. 2022 Phys. Rev. C {\bf 106} 034606

\end{thebibliography}
\end{document}